\documentstyle[12pt,epsfig]{article}

\def\doublespacing{\parskip 5pt plus 1pt
                   \baselineskip 25pt
                   \lineskip 13pt
                   \normallineskip 13pt}

\title{\bf A tracking algorithm for the stable spin polarization field
  in storage rings using stroboscopic averaging}

\author{K.\ Heinemann and G.\ H.\
  Hoffst\"atter\thanks{heineman@mint1.desy.de, hoff@desy.de}
  \\ Deutsches Elektronen--Synchrotron DESY \\ Notkestr.  85,\ 22603 \
  Hamburg}

\begin{document}

\maketitle

\begin{abstract}
  Polarized protons have never been accelerated to more than about
  $25$GeV.  To achieve polarized proton beams in RHIC (250GeV), HERA
  (820GeV), and the TEVATRON (900GeV), ideas and techniques new to
  accelerator physics are needed.  In this publication we will stress
  an important aspect of very high energy polarized proton beams,
  namely the fact that the equilibrium polarization direction can vary
  substantially across the beam in the interaction region of a high
  energy experiment when no countermeasure is taken.  Such a
  divergence of the polarization direction would not only diminish the
  average polarization available to the particle physics experiment,
  but it would also make the polarization involved in each collision
  analyzed in a detector strongly dependent on the phase space
  position of the interacting particle.  In order to analyze and
  compensate this effect, methods for computing the equilibrium
  polarization direction are needed.  In this paper we introduce the
  method of stroboscopic averaging, which computes this direction in a
  very efficient way.  Since only tracking data is needed, our method
  can be implemented easily in existing spin tracking programs.
  Several examples demonstrate the importance of the spin divergence
  and the applicability of stroboscopic averaging.
\end{abstract}

\newpage

\doublespacing

\tableofcontents

\section*{Introduction}
\addcontentsline{toc}{section}{Introduction}

In order to maximize the number of collisions of stored particles in a
storage ring system one tries to maximize the total number of
particles in the bunches and tries to minimize the emittances so that
the particle distribution across phase space is narrow and the phase
space density is high. At equilibrium the phase space distribution
does not change in time and is therefore periodic in the machine
azimuth.

If, in addition, the beam is spin polarized, one requires that the
polarization is high.  As first emphasized by D. Barber
\cite{barber95}, for energies of the order of 1TeV, a fundamental
limitation to the polarization of particle beams becomes important.
To put our work in context we repeat the arguments here.

Spins traveling with particles in electromagnetic fields precess
according to the Thomas--Bargmann--Michel--Telegdi equation (T--BMT)
to be discussed below. The guide fields in storage rings are produced
by dipole and quadrupole magnets. The dipole fields constrain the
particles to almost circular orbits and the quadrupole fields focus
the beam, thus ensuring that the particles do not drift too far away
from the central orbit.

In horizontal dipoles, spins precess only around the vertical field
direction.  The quadrupoles have vertical and horizontal fields and
additionally cause the spins to precess away from the vertical
direction. The strength of the spin precession and the precession axis
in machine magnets depends on the trajectory and the energy of the
particle.  Thus in one turn around the ring the effective precession
axis can deviate from the vertical and will depend on the initial
position of the particle in six dimensional phase space. From this it
is clear that if an equilibrium spin distribution exists, i.e. if the
polarization vector at every phase space point is periodic in the
machine azimuth, it will vary across the orbital phase space.  This
field of equilibrium spin directions in phase space does not change
from turn to turn when particles propagate through the accelerator,
although after each turn the particles find themselves at new
positions in phase space.  These directions, which we denote by the
unit vector ${\vec n}({\vec z}, \theta)$, where $\vec z$ denotes the
position in the six dimensional phase space of the beam and $\theta$
is the generalized azimuth, was first introduced by Derbenev and
Kondratenko \cite{derbenev72} in the theory of radiative electron
polarization.  Note that $\vec n(\vec z,\theta)$ is usually not an
eigenvector of the spin transfer matrix at some phase space point
since the spin of a particle changes after one turn around the ring,
but the eigenvector would not change.

Thus once we know this direction ${\vec n}({\vec z}, \theta)$, the
phase space dependent polarization $p(\vec z,\theta)$ in this
direction, and the phase space density function ${\rho}({\vec z},
\theta)$ we have a complete specification of the polarization state of
a beam of spin $1/2$ particles.

The maximizing of the polarization of the ensemble implies two
conditions; the polarization $p(\vec z,\theta)$ at each point in phase
space should be high and the polarization vector $\vec n(\vec
z,\theta)$ at each point should be almost parallel to the average
polarization vector of the beam.

According to the T--BMT equation, the rate of spin precession is
roughly proportional to $a\gamma$ where $a = (g-2)/2$ is the anomalous
part of the spin $g$ factor and $\gamma$ is the Lorentz factor.
At very high energy, as for example in the HERA proton ring
\cite{barber95}, it could happen that on average ${\vec n}({\vec z},
\theta)$ deviates by tens of degrees from the phase space average of
$\vec n$.  Thus even if each point in phase space were $100\%$
polarized the average polarization could be much smaller than $100\%$.
Clearly it is very important to have accurate and efficient methods
for calculating ${\vec n}({\vec z}, \theta)$ and for ensuring that the
spread of ${\vec n}({\vec z}, \theta)$ is as small as possible.

However, although it has been straightforward to define ${\vec
  n}({\vec z}, \theta)$, this vector is not easy to calculate in
general and much effort has been expended on this topic, but mainly
for electrons at energies up to 46GeV.  Except for the Fourier
expansion formalism introduced in \cite{yokoya92}, all other methods
developed so far are explicitly perturbative, and either do not go to
high enough order \cite{chao81a,chao81b,eidelmann94a} or have problems
with convergence at high order and high energy
\cite{mane87b,balandin92}.

In this paper we describe a new method for obtaining ${\vec n}({\vec
  z}, \theta)$.  It is based on multi--turn tracking and the averaging
of the spin viewed stroboscopically from turn to turn.  Since this
innovative approach only requires tracking data, it is fast and very
easy to implement in existing tracking codes.  We will show that the
convergence speed promises rather quick execution when simulating
realistic accelerators.  However, probably the main advantage over
other methods is the fact that stroboscopic averaging does not have an
inherent problem with either orbit or spin orbit resonances due to its
non-perturbative nature.  This allows the behavior of the periodic spin
solution close to resonances to be analyzed.

\section{The Spin--Orbit System}

The motion of the spin of particles traveling in electromagnetic
fields is governed by the equations of motion

\begin{eqnarray}
  \frac{d\vec z}{d\theta}(\theta) &=&
  \vec v(\vec z(\theta),\theta ) \ ,
  \label{eq:orbdgl}\\
  \frac{d\vec s}{d\theta}(\theta) &=&
  \vec\Omega(\vec z(\theta),\theta )\times
  \vec s(\theta) \ .
  \label{eq:spindgl}
\end{eqnarray}

Here $\theta$ is an independent variable parametrizing the $d$
dimensional particle phase space trajectory $\vec z(\theta)$ and the
spin trajectory $\vec s(\theta)$.  In circular accelerators $\theta$
is the azimuth. The rest frame spin vector $\vec s$ has $3$
components and we normally deal with orbital phase space vectors
$\vec z$ which have $6$ components.  In accelerator physics these
components are usually the positions and momenta of a particle
combined with its energy and the time of flight.

We are neglecting the Stern--Gerlach forces since they are very small in
comparison with the Lorentz force. Equation (\ref{eq:spindgl}) is the
Thomas--BMT equation along an orbit parametrized by $\vec z(\theta)$
\cite{thomas27,bargmann59}. Because we deal with a circular accelerator
at fixed energy, $\vec v(\vec z,\theta)$ and $\vec\Omega(\vec
z,\theta)$, which depend on the guide fields, are periodic in $\theta$
with period $2\pi$ corresponding to the circumference of the ring. Due
to the precession equation (\ref{eq:spindgl}) the length of the spin
vector $\vec s$ does not change along the azimuth.

The dynamical system (\ref{eq:orbdgl},\ref{eq:spindgl}) allows to
formulate the following partial differential equation for the
evolution of a {\em field} $\vec f(\vec z,\theta)$:
\begin{equation}
  \frac{d \vec f}{d\theta} = \frac{\partial\vec f}{\partial\theta}
  + \sum_{j=1}^d v_j\frac{\partial\vec f}{\partial z_j} =
  \vec\Omega\times\vec f \ ,
\label{eq:fdgl}
\end{equation}
where the three components of the solution $\vec f$ depend on $\vec z$
and $\theta$.  In our applications $\vec f$ will describe the
propagation of a spin distribution associated with a particle beam
and this physical interpretation can be adopted because of the
following. A solution $\vec f(\vec z,\theta)$ to equation
(\ref{eq:fdgl}) can be
found by specifying an arbitrary $\vec f(\vec z,\theta_0)$ at initial
azimuth $\theta_0$ and propagating it to $\theta$ by integrating
equations (\ref{eq:orbdgl}) and (\ref{eq:spindgl}). In fact
\begin{equation}
  \vec s(\theta) = \vec f(\vec z(\theta),\theta )
        \label{eq:sf}
\end{equation}
solves equation (\ref{eq:spindgl}) if $\vec f$ solves equation
(\ref{eq:fdgl}).  We say that $\vec f$ is normalized if
\begin{equation}
  |\vec f| = \sqrt{f_1^2+f_2^2+f_3^2} =1 \ .
\end{equation}
We call every normalized solution of equation (\ref{eq:fdgl}) a `spin
field'.

The `$\vec n$-axis' introduced in the Introduction is a special spin
field which is periodic in $\theta$ with period $2\pi$
\cite{derbenev72}:
\begin{equation}
  \vec n(\vec z,\theta+2 \pi) = \vec n(\vec z,\theta) \ .
\label{eq:period}
\end{equation}
Since equation (\ref{eq:spindgl}) represents a pure rotation the
propagation of the spin vector can be described by a $3\times 3$
orthogonal matrix. We denote this rotation matrix which propagates
initial spins $\vec s(\theta_0)$ along a given orbit trajectory $\vec
z(\theta)$ by $\underline R(\vec z(\theta_0),\theta,\theta_0)$ so that
\begin{equation}
  \vec s(\theta) = \underline R(\vec z(\theta_0),\theta,
  \theta_0)\cdot\vec s(\theta_0) \ .
\end{equation}
Because for a spin field $\vec f$ the spin trajectory (\ref{eq:sf})
solves
equation (\ref{eq:spindgl}), we get
\begin{equation}  \vec f(\vec z(\theta),\theta) = \underline
  R(\vec z(\theta_0),\theta,\theta_0)\cdot \vec
  f(\vec z(\theta_0),\theta_0) \ .
\label{eq:ftrans1}
\end{equation}
If $\vec f(\vec z,\theta)$ is a spin field, then
$\vec f(\vec z,\theta+2\pi m)$ is a spin field ($m$= integer).  This
follows from equation (\ref{eq:fdgl}) because $\vec v$ and
$\vec\Omega$
are periodic in $\theta$. Thus equation (\ref{eq:ftrans1}) generalizes
to
\begin{equation}
  \vec f(\vec z(\theta),\theta+2\pi m) = \underline
  R(\vec z(\theta_0),\theta,\theta_0)\cdot \vec
  f(\vec z(\theta_0),\theta_0+2\pi m) \ .  \ \ (m={\rm
    integer})
\label{eq:ftransm}
\end{equation}
Since an $\vec n$-axis is a periodic spin field, we observe
by (\ref{eq:period}) and (\ref{eq:ftransm}) that
\begin{equation}
  \vec n(\vec z(\theta+2\pi),\theta) = \underline
  R(\vec z(\theta),\theta+2\pi,\theta)\cdot\vec
  n(\vec z(\theta),\theta) \ .
\label{eq:nperiod}
\end{equation}
Alternatively this equation can be used for defining the $\vec n$-axis
\cite{yokoya87}. The matrix \\ $\underline R(\vec
z(\theta),\theta+2\pi,\theta)$ is called the one turn spin transfer
matrix for the trajectory $\vec z(\theta)$.

In the special case where the orbital motion is determined by a
Hamiltonian we have
\begin{equation}
  \vec v(\vec z,\theta)= \lbrace
  \vec z,H_{orb}(\vec z,\theta)\rbrace \ ,
\end{equation}
where $H_{orb}$ denotes the orbital Hamiltonian.  Furthermore in this
case one can define for the whole spin--orbit system a Hamiltonian given
by \cite{derbenev73}
\begin{equation}
  H(\vec z,\vec s,\theta) = H_{orb}(\vec z,\theta) +
  H_{spin}(\vec z,\vec s,\theta) \ ,
\end{equation}
where
\begin{equation}
  H_{spin}(\vec z,\vec s,\theta) =
  \vec s\cdot\vec\Omega(\vec z,\theta) \ .
\end{equation}
The Poisson brackets of this Hamiltonian lead to equations
(\ref{eq:orbdgl}) and (\ref{eq:spindgl}) if the Stern--Gerlach forces
are neglected \cite{barber94b}.

\section{Construction of Periodic Spin Fields by Stroboscopic
Averaging}

To find solutions of equation (\ref{eq:fdgl}) which are periodic in
$\theta$ by our new method, one first constructs an arbitrary spin
field $\vec f$.  One then constructs the following `stroboscopic
average' of $\vec f$:
\begin{equation}
  <\vec f>(\vec z,\theta) = \lim_{n\rightarrow \infty}[
  \frac{1}{n+1}\sum_{m=0}^{n} \vec f(\vec z,\theta+2\pi m)]\ .
\label{eq:faver}
\end{equation}
Since the convergence and differentiability of the sequence in
(\ref{eq:faver}) can in general not be guaranteed, the limit is only
taken formally. The problem of the convergence will be analyzed in
more detail in section
\ref{ssec:conv}.
From (\ref{eq:faver}) it follows that $<\vec f>$ is formally periodic
in $\theta$. Moreover, because (\ref{eq:fdgl}) is a linear equation
and because $\vec v$ and $\vec\Omega$ are periodic in $\theta$ we
observe for any spin field $\vec f$ that $<\vec f>$ is formally also a
solution of equation (\ref{eq:fdgl}). Hence we come to a first
conclusion:
\begin{itemize}
\item If $\vec f$ is a spin field, then $<\vec f>$ is a solution of
equation (\ref{eq:fdgl}) which is periodic in $\theta$.
\end{itemize}
If for $\vec f$ the stroboscopic average $<\vec f>$ vanishes nowhere
in the $d+1$ dimensional space, then we define
\begin{equation}
  <\vec f>_{norm} = <\vec f>/|<\vec f>| \ .
\end{equation}
In general $<\vec f>$ is not normalized, but the modulus of $<\vec f>$
is conserved and $<\vec f>_{norm}$ is a spin field which is
periodic in $\theta$. Hence we come to the following second conclusion
of this section:
\begin{itemize}
\item If $\vec f$ is a spin field with the property that $<\vec f>$
  vanishes nowhere in the $d+1$ dimensional space, then $<\vec
  f>_{norm}$ has all the properties of an $\vec n$-axis.
\end{itemize}
This result shows that an $\vec n$-axis can be obtained from a spin
field $\vec f$ for which $<\vec f>$ vanishes nowhere.
In the next section we will derive a tracking
algorithm based on this.

One practical choice of $\vec f$ is characterized for all $\vec z$ by
\begin{equation}
  \vec f(\vec z,\theta_0) = \vec n_0(\theta_0)\ ,
\label{eq:fnzero}
\end{equation}
where $\vec n_0(\theta)$ denotes the so called closed orbit spin
axis defined by
\begin{equation}
  \vec n_0(\theta) = \underline R(\vec
  z_{c.o.}(\theta),\theta+2\pi,\theta) \cdot\vec n_0(\theta) \ ,
\end{equation}
where $\vec z_{c.o.}(\theta)$ is the closed orbit.

\section{The Tracking Algorithm for the $\vec n$-Axis
  using Stroboscopic Averaging}
\label{sec:algo}

In this section we develop a tracking algorithm which provides an
efficient way to evaluate an $\vec n$-axis at $\vec z=\vec z_0,
\theta =\theta_0$.  Choosing a spin field $\vec f$ and replacing
$\theta_0$ by $\theta_0-2\pi m$ in equation (\ref{eq:ftransm}) we get
for every integer $m$
\begin{equation}
  \vec f(\vec z(\theta),\theta+2\pi m) = \underline R(\vec
  z(\theta_0-2\pi m),\theta,\theta_0- 2\pi m)\cdot \vec
  f(\vec z(\theta_0-2\pi m),\theta_0)\ .
\end{equation}
If we choose an orbit with $\vec z(\theta_0)=\vec z_0$, then
inserting this into equation (\ref{eq:faver}) results in
\begin{equation}
  <\vec f>(\vec z_0,\theta_0) = \lim_{n\rightarrow \infty}[
  \frac{1}{n+1}\sum_{m=0}^{n} \underline R(\vec z(\theta_0-2\pi m),
  \theta_0,\theta_0-2\pi m)\cdot \vec f(\vec z(\theta_0-2\pi m),
  \theta_0)]\ .
\label{eq:aver}
\end{equation}
Normalization of $<\vec f>$ yields an $\vec n$-axis at
$(\vec z_0,\theta_0)$.

To apply the tracking algorithm, the infinite sum involved in the
stroboscopic average (\ref{eq:aver}) is replaced by a finite sum of
$N+1$ terms so that we approximate
\begin{eqnarray}
  <\vec f>(\vec z_0,\theta_0) &\approx& <\vec f>_N(\vec z_0,\theta_0)
  = \frac{1}{N+1} \sum_{m=0}^{N} \vec f(\vec z_0,\theta_0+2\pi m)
  \nonumber\\ &=& \frac{1}{N+1}\sum_{m=0}^{N} \underline R(\vec
  z(\theta_0-2\pi m), \theta_0,\theta_0-2\pi m)\cdot \vec
  f(\vec z(\theta_0-2\pi m),\theta_0)\ ,
\label{eq:favern}
\end{eqnarray}
which yields the following approximation of the $\vec n$-axis
\begin{equation}
\vec n(\vec z_0,\theta_0) \approx
   \frac{    <\vec f>_N(\vec z_0,\theta_0) }
     { |<\vec f>_N (\vec z_0,\theta_0)| }\ .
\label{eq:nfnaver}
\end{equation}
The stroboscopic average $<\vec f>_N$ in equation (\ref{eq:favern})
has a very simple physical interpretation which illustrates its
practical importance. If a particle beam is approximated by a phase
space density, disregarding its discrete structure, then we can
associate a spin field $\vec f(\vec z,\theta_0)$ with the particle
beam at the azimuth $\theta_0$. If one installs a point like
`gedanken' polarimeter at a phase space point $\vec z_0=\vec
z(\theta_0)$ and azimuth $\theta_0$, then this polarimeter initially
measures $\vec f(\vec z_0,\theta_0)$. When the particle beam passes
the azimuth $\theta_0$ after one turn around the ring, the polarimeter
measures $$\underline R(\vec
z(\theta_0-2\pi),\theta_0,\theta_0-2\pi)\cdot\vec f(\vec
z(\theta_0-2\pi),\theta_0)=\vec f(\vec z_0,\theta_0+2\pi)\ .$$ After
the beam has traveled around the storage ring $N$ times and the
polarization has been measured whenever the beam passed the `gedanken'
polarimeter, one only has to average over the different measurements
in order to obtain $<\vec f>_N$. If the particles of a beam are
polarized parallel to $\vec n(\vec z,\theta_0)$ at every phase space
point, then the spin field of the beam is invariant from turn to turn
due to the periodicity property in equation (\ref{eq:period}). But in
addition, even for beams which are not polarized parallel to $\vec n$,
we see that the polarization observed at a phase space point $\vec z$
and azimuth $\theta_0$ is still parallel to $\vec n(\vec z,\theta_0)$,
if one averages over many measurements taken when the beam has passed
the azimuth $\theta_0$.

For the special
choice $\vec f(\vec z,\theta_0)=\vec n_0(\theta_0)$ we can simplify
$<\vec f>_N$ to
\begin{equation}
  <\vec f>_N(\vec z_0,\theta_0) = \frac{1}{N+1} \sum_{m=0}^{N}
  \underline R(\vec z(\theta_0-2\pi m), \theta_0,\theta_0-2\pi
  m)\cdot\vec n_0(\theta_0) \ .\label{eq:avfnzero}
\end{equation}
Equations (\ref{eq:favern}) and (\ref{eq:nfnaver}) define an algorithm
for obtaining an $\vec n$-axis. We see that the only information
needed from tracking is the set of the $N+1$ phase space points $\vec
z(\theta_0), \vec z(\theta_0-2\pi),\ldots,\vec z(\theta_0-2\pi N)$ and
the $N$ matrices $\underline R(\vec z(\theta_0-2\pi),\theta_0,
\theta_0-2\pi),\underline R(\vec z(\theta_0-4\pi),
\theta_0,\theta_0-4\pi),\ldots, \underline R(\vec z(\theta_0-2\pi
N),\theta_0, \theta_0-2\pi N)$. Each matrix is a product of one turn
spin transfer matrices $\underline R(\vec z,\theta_0+2\pi,\theta_0)$.
This means that one tracks along the orbit $\vec z(\theta)$ to obtain
the spin transfer matrix $\underline R(\vec
z(\theta),\theta_0,\theta)$ and stores it at the $N$ instants, where
$\theta=\theta_0-2\pi N,\ldots,
\theta=\theta_0-4\pi,\theta=\theta_0-2\pi$.  The function $\vec f(\vec
z,\theta_0)$ is chosen independently of the tracking results (for
example one can take the choice $\vec f(\vec z,\theta_0)=\vec
n_0(\theta_0)$ of equation (\ref{eq:fnzero}) and (\ref{eq:avfnzero})).

The following two kinds of pathologies can occur:
\begin{itemize}
\item The $\vec n$-axis is not unique: if the proposed algorithm
  converges, then the result could depend on the choice of $\vec
  f(\vec z,\theta_0)$.
\item The stroboscopic average
  $<\vec f>$ vanishes for $N\rightarrow+\infty$ or the sequence in
  equation (\ref{eq:faver}) does not converge.
\end{itemize}
Both pathologies can be studied with the algorithm. The first
situation occurs for systems on spin--orbit resonances
\cite{yokoya86}.  In all examples studied so far, the stroboscopic
average seems to converge, implying the existence of an $\vec n$-axis.
In the second situation, the point like polarimeter at $\vec z_0$
mentioned above monitors an averaged polarization which either
vanishes or fluctuates indefinitely.

\section{Efficient Implementation Only Using One Turn Information}

In the previously outlined formalism for evaluating an $\vec n$-axis
by stroboscopic averaging it became apparent that only knowledge about
one turn spin transfer matrices is required.  One can therefore
reformulate the algorithm of section \ref{sec:algo} in terms of one
turn maps which are to be taken at a fixed but arbitrary azimuth value
$\theta_0$ and thereby obtain a more practical algorithm.  Thus we
introduce the one turn orbit transfer map $\vec M$ which maps initial
coordinates $\vec z_i$ into final coordinates $\vec z_f=\vec M(\vec
z_i)$.  Then in our notation we have $\vec z(\theta_0+2\pi) = \vec
M(\vec z(\theta_0))$.  To describe the transport of particles with
spins $\vec s$, we write for simplification the one turn spin transfer
matrix $\underline R(\vec z,\theta_0+2\pi,\theta_0)$ as $\underline
R(\vec z)$ so that we have $\vec s_f=\underline R(\vec z_i)\cdot\vec
s_i$.  All other quantities which depend on $\theta$ are taken at the
specified azimuth $\theta_0$.  For simplification, this azimuth is not
indicated in the following.  As already mentioned, equation
(\ref{eq:nperiod}) can be used to define $\vec n$-axes at $\theta_0$.
This condition will be called the periodicity condition and now reads
\begin{equation}
  \underline R(\vec z)\cdot\vec n(\vec z) = \vec n(\vec M(\vec z))\ .
\label{eq:per}
\end{equation}

\subsection{Recipe}
\label{ssec:rec1}

To illustrate the process of evaluating an $\vec n$-axis at $\vec z_0$
and $\theta_0$ in the case of linear orbit motion, we establish a
recipe.
\begin{enumerate}
\item Compute the linearized one turn phase space transfer map $\vec
  z_f=\underline M\cdot\vec z_i$.
\item Define the set of $N+1$ phase space points
  \begin{equation}
    C=\{\vec c_j=(\underline M^{-1})^j\cdot\vec
    z_0|j\in\{0,\ldots,N\}\}\ .
  \end{equation}
\item Compute the rotation matrix $\underline R(\vec z_{c.o.})$ which
  describes the spin motion for particles on the closed orbit $\vec
  z_{c.o.}(\theta)$ and extract the corresponding rotation vector
  $\vec n_0$.  This is the periodic spin solution for particles on the
  closed orbit.
\item Starting with a spin parallel to $\vec n_0$ at every phase
  space point in $C$, track until the phase space point $\vec z_0$ is
  reached.  For a given $j$ this requires tracking $j$ turns starting at
  $\vec c_j$.
\item Define the set of spin tracking results as
\begin{equation}
  B=\{\vec b_0(\vec z_0)=\vec n_0, \vec b_j(\vec z_0)=\underline
  R(\vec c_1)\cdot\ldots\cdot\underline R(\vec c_j)\cdot\vec n_0 |
  j\in\{1,\ldots,N\}\}\ .
\end{equation}
\item Define the sum of the elements in $B$ as $\vec s_N(\vec z_0) =
  \frac{1}{N+1}\sum_{j=0}^N \vec b_j(\vec z_0)$ and for $|\vec s_N|\ne
  0$ define $\vec n_N=\vec s_N/|\vec s_N|$.
\end{enumerate}
The vector $\vec s_N(\vec z_0)$ is equivalent to $<\vec f>_N(\vec z_0,
\theta_0)$ in section
\ref{sec:algo}, if the initial distribution of spins is given by $\vec
n_0$ as in equation (\ref{eq:avfnzero}).

\subsection{Convergence Speed}
\label{ssec:conv}

It will now be shown that if the angle between $\vec n_0$ and $\vec
b_j(\vec z_0)$ is smaller than some positive number $\xi<\pi/2$ for
all $j\in\{1,\ldots,N+1\}$, then $\vec n_N$ satisfies the periodicity
condition (\ref{eq:per}) for the $\vec n$-axis up to an error which is
smaller or equal to $2\sec(\xi/2)\tan(\xi)/(N+1)$.  Since evaluating
$B$ by the recipe of section \ref{ssec:rec1} requires tracking
$T=(N+1)N/2$ turns, the accuracy is bounded by
$\sqrt{2/T}\sec(\xi/2)\tan(\xi)$.  This slow convergence with the
square root of $T$ is a very serious limitation and in the next
section we will demonstrate how the convergence can be considerably
improved.

The proof of this convergence property goes along the following lines.
The average $\vec s_N$ has been defined by
\begin{equation}
  \vec s_N(\vec z_0) = \frac{1}{N+1}\sum_{j=0}^N \prod_{k=1}^j
  \underline R(\vec c_k)\cdot\vec n_0\ .
\label{eq:sn}
\end{equation}
Here we adopt the convention $\prod_{k=1}^0\underline R(\vec c_k) =
\prod_{k=1}^0\underline R(\vec c_{k-1}) = \underline 1$ and
$\prod_{k=1}^j\underline R(\vec c_k)$ is taken to mean the following
order of multiplication: $\underline R(\vec
c_1)\cdot\ldots\cdot\underline R(\vec c_j)$.  To check how well $\vec
s_N$ satisfies the periodicity conditions (\ref{eq:per}) of the $\vec
n$-axis we calculate
\begin{eqnarray}
  \vec s_N(\underline M\cdot\vec z_0) &=& \frac{1}{N+1}\sum_{j=0}^N
  \prod_{k=1}^j \underline R(\vec c_{k-1})\cdot\vec n_0 \nonumber\\ 
  &=& \frac{1}{N+1}(\vec n_0+\sum_{j=0}^{N-1} \prod_{k=0}^j \underline
  R(\vec c_{k})\cdot\vec n_0)\ ,\\ \underline R(\vec z_0)\cdot\vec
  s_N(\vec z_0) &=& \frac{1}{N+1}\sum_{j=0}^N \prod_{k=0}^j \underline
  R(\vec c_{k})\cdot\vec n_0\ ,\\ \underline R(\vec z_0)\cdot\vec
  s_N(\vec z_0) - \vec s_N(\underline M\cdot\vec z_0) &=&
  \frac{1}{N+1}(\underline R(\vec z_0)\cdot\vec b_N(\vec z_0)-\vec
  n_0)\nonumber\\ &=& \frac{1}{N+1}(\vec b_{N+1}(\underline M\cdot\vec
  z_0)-\vec n_0).
\end{eqnarray}
\begin{figure}[htbp]
\begin{center}
  \setlength{\unitlength}{1cm}
  \psfig{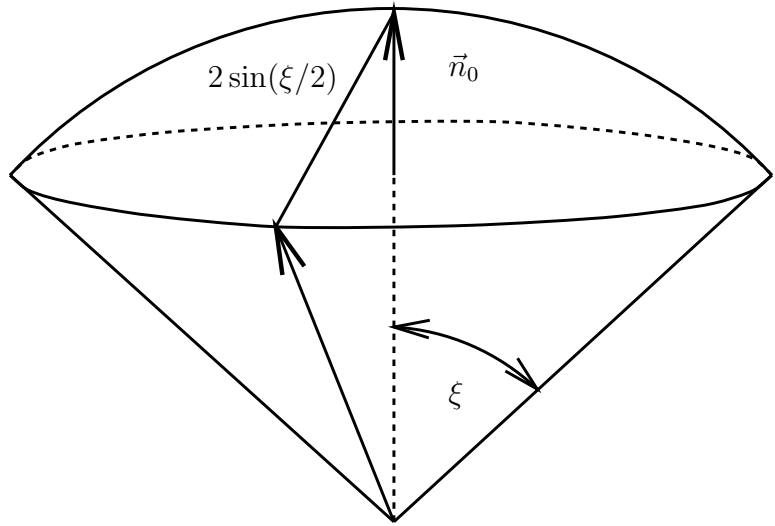}
\begin{picture}(0,0)(0,0)
  \put(-4.5,-5.35) {$\xi$} \put(-4.5,-1.0) {$\vec n_0$}
  \put(-7.7,-1.15) {$2\sin(\xi/2)$}
\end{picture}
\end{center}
\caption{Estimation of convergence speed}
\label{fg:convergence}
\end{figure}

The length $|\vec b_{N+1}(\underline M\cdot\vec z_0)-\vec n_0|$ is
smaller than $2\sin(\xi/2)$ as shown in figure \ref{fg:convergence}.
The length of $\vec s_N$ is at least $\cos(\xi)$; and here it becomes
essential that there is limit of $\pi/2$ on the angle $\xi$.  This
information is sufficient to establish the following inequality
\begin{eqnarray}
  |\underline R(\vec z_0)\cdot\vec n_N(\vec z_0)-\vec n_N(\underline
  M\cdot\vec z_0)| &=& |\underline R(\vec z_0)\cdot\frac{\vec s_N(\vec
    z_0)}{|\vec s_N(\vec z_0)|} -\frac{\vec s_N(\underline M\cdot\vec
    z_0)}{|\vec s_N(\underline M\cdot\vec z_0)|}| \nonumber\\ &=&
  \frac{1}{|\vec s_N(\vec z_0)|} |\underline R(\vec z_0)\cdot\vec
  s_N(\vec z_0)-\vec s_N(\underline M\cdot\vec z_0) \nonumber\\ 
  &&+\frac{|\vec s_N(\underline M\cdot\vec z_0)|-|\vec s_N(\vec z_0)|}
  {|\vec s_N(\underline M\cdot\vec z_0)|}\vec s_N(\underline
  M\cdot\vec z_0)| \nonumber\\ &\leq& \frac{1}{|\vec s_N(\vec z_0)|}(
  |\underline R(\vec z_0)\cdot\vec s_N(\vec z_0)-\vec s_N(\underline
  M\cdot\vec z_0)| \nonumber\\ &&+||\underline R(\vec z_0)\vec
  s_N(\vec z_0)|-|\vec s_N(\underline M\cdot\vec z_0)||) \nonumber\\ 
  &\leq& \frac{2}{|\vec s_N(\vec z_0)|} |\underline R(\vec
  z_0)\cdot\vec s_N(\vec z_0)-\vec s_N(\underline M\cdot\vec z_0)|
  \nonumber\\ &\leq& \frac{4\sin(\xi/2)}{(N+1)\cos(\xi)}\ .
\end{eqnarray}
The error by which the vector $\vec n_N(\vec z_0)$ violates the
periodicity condition (\ref{eq:per}) of the $\vec n$-axis is therefore
smaller than $2\sec(\xi/2)\tan(\xi)/(N+1)$ and converges to $0$ for
large $N$.

If one can prove the existence of a suitable number $\xi<\pi/2$ for
some spin transport system, one has proven the existence of functions
$\vec n_N$ for this system which satisfy the defining equation for the
$\vec n$-axis to arbitrary precision.  Since, however, these
functions $\vec n_N$ do not necessarily converge, this does not prove
the existence of an $\vec n$-axis for such a system.

If the orbit motion can be described in terms of action--angle
variables, as is always the case for linear motion, and the orbital
angle advances for one turn ($2\pi$ times the orbit tunes) are not in
resonance, then two important conclusions about this tracking
algorithm with $\xi<\pi/2$ can be drawn.
\begin{enumerate}
\item If an $\vec n$-axis $\vec n(\vec z)$ exists, then the sequence
  $\vec n_N$ converges to $\vec n(\vec z)$ linearly in $1/N$.
\item If an $\vec n$-axis exists and the spin rotation angle in one
  turn is not a linear combination of orbit phase advances modulo
  $2\pi$, then the $\vec n$-axis is unique up to a sign.
\end{enumerate}

The proof is adapted from \cite{yokoya92,yokoya86,barber92}.  The
first step will be to show how to define a spin rotation angle which
depends only on orbital action variables.  We assume that an $\vec
n$-axis exists and introduce two unit vectors $\vec u_1(\vec z)$ and
$\vec u_2(\vec z)$ to create a right handed coordinate system $(\vec
n,\vec u_1,\vec u_2)$.  The vectors $\vec u_1$ and $\vec u_2$ are
therefore defined up to a rotation around the $\vec n$-axis by an
arbitrary phase space dependent angle $\phi(\vec z)$.  We express the
spin vectors $\vec s$ in terms of this coordinate system by $\vec
s=s_1\vec n+s_2\vec u_1+s_3\vec u_2$.  The coefficient $s_1$ does not
change during the particle motion around the ring since the particle
transfer matrix $\underline R(\vec z)$ is orthogonal and ensures that
$(\vec s\cdot\vec n)$ is invariant.  The spin motion described by the
$\underline R(\vec z)$ matrix is therefore simply a rotation around
the $\vec n$-axis by a phase space dependent angle $\nu(\vec z)$.
\begin{equation}
\left(\begin{array}{c}s_{f1}\\s_{f2}\\s_{f3}\end{array}\right)
        = \left(\begin{array}{rrr}1&0&0\\
                0& \cos(\nu(\vec z))&\sin(\nu(\vec z))\\
              0&-\sin(\nu(\vec z))&\cos(\nu(\vec z))\end{array}\right)
\left(\begin{array}{c}s_{i1}\\s_{i2}\\s_{i3}\end{array}\right)\ .
\end{equation}

If we now introduce the complex quantity $\hat s = e^{-i\phi(\vec
  z)}(s_2+is_3)$ where $\phi(\vec z)$ is the arbitrary angle, then the
spin transport is described by
\begin{eqnarray}
  s_{f2}+is_{f3}&=&e^{-i\nu(\vec z)}(s_{i2}+is_{i3})\ , \\
  e^{i\phi(\vec M(\vec z))}\hat s_f &=& e^{i(-\nu(\vec z)+\phi(\vec
    z))}\hat s_i\ .
\end{eqnarray}
Now we introduce orbital action--angle variables $\vec J$ and
$\vec\Phi$ as well as the angle advances $\vec Q$ for one turn around
the accelerator.  Note that the symbol $\vec Q$ is $2\pi$ times the
orbital tunes.  In these variables the one turn transport is
characterized by $\vec J_f=\vec J_i$ and $\vec\Phi_f=\vec\Phi_i+\vec
Q$.  Since the actions remain invariant during the particle motion, we
use the symbols $\nu_{\vec J}(\vec\Phi)$ and $\phi_{\vec J}(\vec\Phi)$
to indicate the spin rotation angle and the free phase of the
coordinate system for fixed actions $\vec J$
\begin{equation}
  \hat s_f = e^{i(-\nu_{\vec J}(\vec\Phi)+\phi_{\vec
      J}(\vec\Phi)-\phi_{\vec J}(\vec\Phi+\vec Q))}\hat s_i\ .
\label{eq:expon}
\end{equation}
We now show how $\phi_{\vec J}(\vec\Phi)$ can be chosen so that the
spin motion characterized by the exponent becomes simplified.  As with
any function of phase space, the rotation $e^{i\phi_{\vec
    J}(\vec\Phi)}$ is $2\pi$ periodic in all components $\Phi_j$.
Therefore, the rotation angle $\phi_{\vec J}(\vec\Phi)$ has a periodic
contribution $\phi_{\circ\vec J}(\vec\Phi)$ and a linear contribution
in the phases
\begin{equation}
  \phi_{\vec J}(\vec\Phi) = \phi_{\circ\vec J}(\vec\Phi) + \vec
  j\cdot\vec\Phi
\end{equation}
with some vector $\vec j$ that has integer components.  The phase
space dependent spin rotation $e^{i\nu_{\vec J}(\vec\Phi)}$ is also a
periodic function of phase space.  But since on the closed orbit
($\vec J=0$) the spin motion does not depend on $\vec\Phi$, therefore
$\nu_{\vec J}(\vec\Phi)$ only has a periodic component and no
component linear in $\vec\Phi$.

If the orbital angle advances $\vec Q$ are not in resonance, then
$\phi_{\vec J}(\vec\Phi)$ can be chosen to eliminate the phase
dependence of the exponent in equation (\ref{eq:expon}) completely.
This can be seen by Fourier transformation of the periodic functions
$\nu_{\vec J}(\vec\Phi)$ and $\phi_{\circ\vec J}(\vec\Phi)$ leading to
the requirement
\begin{equation}
  \breve\nu_{\vec J}(\vec k)=\breve\phi_{\circ\vec J}(\vec
  k)(1-e^{i\vec k\cdot\vec Q})
\end{equation}
for the Fourier coefficients $\breve\nu_{\vec J}(\vec k)$ and
$\breve\phi_{\circ\vec J}(\vec k)$.  Therefore, $\breve\phi_{\circ\vec
  J}(\vec k)$ can be chosen to eliminate all Fourier coefficients
except for $\vec k=0$.  With this choice the exponent reduces to
$\breve\nu_{\vec J}(0)+\vec j\cdot\vec Q$.  The vector $\vec j$ can in
general not be used to simplify this expression and we therefore
usually choose $\vec j=0$.  With $\nu(\vec J)=\breve\nu_{\vec J}(0)$, the
spin rotation of equation (\ref{eq:expon}) then simplifies to
\begin{equation}
\hat s_f=e^{-i\nu(\vec J)}\hat s_i  \ .
\end{equation}
We have now achieved the goal of constructing a spin rotation
depending only on orbital actions.  It is interesting to note that if
for amplitudes $\vec J$ the integer coefficients of $\vec j$ can be
chosen so that
\begin{equation}
  \nu(\vec J)+\vec j\cdot\vec Q = 0 {\rm \ mod\ } 2\pi\ .
\label{eq:res}
\end{equation}
one can eliminate the spin rotation completely.  Here we only analyze
the case when this resonance condition is not satisfied.

The stroboscopic average is now performed by the recipe of section
\ref{ssec:rec1} but this time in the coordinate system $(\vec n,\vec
u_1,\vec u_2)$ just constructed.  We first establish the tracking
points $\vec c_j$ and note that $\nu=\nu(\vec c_j)$ is the same for
all tracking points.  In the coordinate system $(\vec n,\vec u_1,\vec
u_2)$, the vector components of the periodic spin $\vec n_0$ on the
closed orbit are not constant but depend on the phase space position
$(\vec J,\vec\Phi)$.  This vector is transported from the phase space
points $\vec c_j$ to $\vec z_0=(\vec\Phi,\vec J)$ by the rotation matrix
\begin{equation}
  \left(\begin{array}{rrr} 1&0 &0 \\ 0& \cos(j\nu)&\sin(j\nu)\\
    0&-\sin(j\nu)&\cos(j\nu)\end{array}\right)\ ,
\end{equation}
leading to the stroboscopic average in this coordinate system
\begin{eqnarray}
  \vec s_N(\vec\Phi,\vec J) &=& \frac{1}{N+1}\sum_{j=0}^N
  \left(\begin{array}{rrr} 1&0 &0 \\ 0& \cos(j\nu)&\sin(j\nu)\\ 
    0&-\sin(j\nu)&\cos(j\nu)\end{array}\right)
    \left(\begin{array}{c}n_{0,1\vec J}(\vec\Phi-j\vec Q)\\n_{0,2\vec
        J}(\vec\Phi-j\vec Q)\\n_{0,3\vec J}(\vec\Phi-j\vec
      Q)\end{array}\right) \ .
\end{eqnarray}
The first component of $\vec s_N$ is
$\sum_{j=0}^Nn_{0,1\vec J}(\vec\Phi-j\vec Q)/(N+1)$, the second and
third components in complex notation are
\begin{equation}
  \hat s_N = s_{N2}+is_{N3} =
  \frac{1}{N+1}\sum_{j=0}^Ne^{-i(j\nu)}\hat n_{0,\vec
    J}(\vec\Phi-j\vec Q)\ ,
\end{equation}
where $\hat n_{0,\vec J}(\vec\Phi)=n_{0,2\vec J}(\vec\Phi)+in_{0,3\vec
  J}(\vec\Phi)$.  In terms of the Fourier components $\breve n_{0,\vec
  J}(\vec k)$ of $\hat n_{0,\vec J}(\vec\Phi)$ one obtains the
inequality
\begin{eqnarray}
  |\hat s_N| &=& |\frac{1}{N+1}\sum_{j=0}^N e^{-ij(\nu+\vec k\cdot\vec
    Q)}\sum_{\vec k}\breve n_{0,\vec J}(\vec k)e^{i\vec
    k\cdot\vec\Phi}|\nonumber\\ &=& \frac{1}{N+1}|\sum_{\vec
    k}\frac{1-e^{-i(N+1)(\nu+\vec k\cdot\vec Q)}}{1-e^{-i(\nu+\vec
        k\cdot\vec Q)}}\breve n_{0,\vec J}(\vec k)e^{i\vec
      k\cdot\vec\Phi}| \nonumber\\ &\leq& \frac{1}{N+1}\sum_{\vec
      k}\frac{2}{|1-e^{-i(\nu+\vec k\cdot\vec Q)}|}|\breve
    n_{0,\vec J}(\vec k)|\ .
\label{eq:sncon}
\end{eqnarray}
Similarly one obtains for the first component of $\vec s_N$
\begin{equation}
  s_{N,1} = \breve n_{0,1\vec J}(0)+\frac{1}{N+1}\sum_{\vec
    k\neq0}\frac{1-e^{-i(N+1)\vec k\cdot\vec Q}}{1-e^{-i\vec
      k\cdot\vec Q}}\breve n_{0,1\vec J}(\vec k)e^{i\vec
    k\cdot\vec\Phi}\ .\label{eq:sonen}
\end{equation}
Since $|\vec s_N|>\cos(\xi)$ and $\xi<\pi/2$, it is guaranteed that
$\vec s_N$ does not converge to $0$.  For large $N,$ $|\vec s_N|$ is
given by $\breve n_{0,1\vec J}$, which therefore does not vanish.
Since the resonance condition (\ref{eq:res}) is not satisfied for any
vector of integers $\vec j$, none of the denominators is zero and
$s_{N,1}$ and $\hat s_N$ converge linearly with $1/N$.  At this point
we impose the condition on $\vec n$ that the sum on the right hand
side of equation (\ref{eq:sncon}) converges.  From the convergence of
$\vec s_N$ the convergence of $\vec n_N$ to $\vec n$ follows from
\begin{eqnarray}
  |\vec n-\vec n_N|^2&=&|\vec n-\frac{\vec s_N}{|\vec s_N|}|^2=
  (1-\frac{s_{N,1}}{|\vec s_N|})^2+(\frac{|\hat s_N|}{|\vec s_N|})^2 =
  2\frac{|\vec s_N|-s_{N,1}}{|\vec s_N|}\nonumber\\ &<& 2\frac{|\vec
    s_N|-s_{N,1}}{|s_{N,1}|}< 2(\frac{|\hat s_{N}|}{|s_{N,1}|})^2\ .
\end{eqnarray}
There is a number $N^*$ such that the absolute value of the $N$
dependent part of $s_{N,1}$ in equation (\ref{eq:sonen}) is smaller
than $\breve n_{0,1\vec J}(0)/2$.  If the number of turns $N$ is
bigger than $N^*$, we conclude
\begin{equation}
  |\vec n_N-\vec n| < 2\sqrt{2}\frac{|\hat s_{N}|}{|n_{0,1\vec J}(0)|}
\end{equation}
which, together with equation (\ref{eq:sncon}), establishes that
$\vec n_N$ convergences linearly with $1/N$.

The second conclusion to be proven is the uniqueness of the $\vec
n$-axis.  In the coordinate system $(\vec n,\vec u_1,\vec u_2)$, the
periodicity condition (\ref{eq:per}) reads
\begin{equation}
\left(\begin{array}{rrr}
1&0         &0        \\
0& \cos(\nu)&\sin(\nu)\\
0&-\sin(\nu)&\cos(\nu)
\end{array}\right)\cdot\vec n(\vec z)=\vec n(\vec M(\vec z))
\label{eq:perioco}
\end{equation}
with the obvious solution $\vec n(\vec z)=(1,0,0)^T$ for all $\vec z$.
If another $\vec n$-axis $\vec n_2(\vec z)$ exists, then $\vec n_2-\vec
n(\vec n\cdot\vec n_2)$ is nonzero at least at one phase space point
and on all iterates of this point which can be reached during particle
motion.  We normalize this difference vector at these phase space
points and write it as $\cos(\alpha(\vec z))\vec u_1+\sin(\alpha(\vec
z))\vec u_2$.  In orbital action--angle variables, the free function
$\vec\alpha_{\vec J}(\vec\Phi)$ again has a periodic contribution and
a linear contribution $\vec j\cdot\vec\Phi$.

In complex notation, the periodicity condition (\ref{eq:perioco}) reads
\begin{equation}
  e^{i(-\nu(\vec J)+\alpha_{\vec J}(\vec\Phi))}=e^{i\alpha_{\vec
      J}(\vec\Phi+\vec Q)}\ .
\label{eq:alph}
\end{equation}
Since $\nu(\vec J)$ does not depend on $\vec\Phi$, the periodic part
of $\alpha$ has to vanish.  The resulting $\alpha_{\vec J}=\vec
j\cdot\vec\Phi$ can only solve the periodicity condition if there is a
vector $\vec j$ with integer coefficients fulfilling the resonance
equation (\ref{eq:res}) and then a second $\vec n$-axis exists.  For
linear orbit motion with phase advances which are not on resonances of
type (\ref{eq:res}), equation (\ref{eq:alph}) does not have a solution
and the $\vec n$-axis is unique.  There is so far no proof for
nonlinear orbits, but nevertheless the method of stroboscopic
averaging can also be used for exploring nonlinear orbit motion.

\subsection{Improved Recipe with Faster Convergence}
\label{ssec:rec2}

In section \ref{ssec:rec1} an algorithm was introduced which converged
with the square root of the number, $T$, of turns tracked.  However, it
is possible to obtain convergence linear in the $T$, when one takes
advantage of the orthogonality of the spin transfer matrix.  To
illustrate this, we again establish a recipe.
\begin{enumerate}
\item Define as before
\begin{equation}
  C=\{\vec c_j=(\underline M^{-1})^j\cdot\vec
  z_0|j\in\{0,\ldots,N\}\}\ .
\end{equation}
\item Define three orthogonal unit vectors
      $\{\vec e^{(1)},\vec e^{(2)},\vec e^{(3)}\}$.
\item Obtain the sets $S_j$ of the three vectors $\vec s_j^{(1)}$,
      $\vec s_j^{(2)}$, and $\vec s_j^{(3)}$ by tracking the
      $\vec e^{(k)}$ for $N-j$ turns
\begin{equation}
  \vec s_j^{(k)} = \underline R(\vec
  c_{j+1})\cdot\ldots\cdot\underline R(\vec c_N)\cdot\vec e^{(k)}\ .
\end{equation}
\item From the set of vectors $S_j$ and the set $S_0$, one can obtain
  the spin transfer matrix from the phase space point $\vec c_j$ to
  $\vec z_0$ denoted by $\underline{\bar R}(\vec c_j)$.  This becomes
  clear when one realizes that $\vec s_0^{(k)} = \underline{\bar
    R}(\vec c_j)\cdot\vec s_j^{(k)}$ for all $k$.  Obtaining these
  $N+1$ transport matrices requires the propagation of 3 spins around
  the circular accelerator for $N$ turns.
\item Now we can compute the set $B = \{\underline{\bar R}(\vec
  c_j)\cdot\vec n_0|j\in\{0,\ldots,N\}\}$.  This is obviously
  identical to the set denoted by $B$ in the previous section.
\item The normalized average of $B$, denoted by $\vec n_N$, can now
  again be computed as mentioned above and solves the defining
  equation for an $\vec n$-axis up to the small error discussed
                             previously.
\end{enumerate}

In this approach one only has to track three initial spin directions
over $N$ turns, leading to $T=3N$.  The error is therefore bounded by
$6\sec(\xi/2)\tan(\xi)/T$.  This implies convergence linear in the
number $T$ of turns tracked.  The following example illustrates the
                              speed of this
method:  when the angle $\xi$ happens to be $45^\circ$, and we require
an accuracy at the $10^{-3}$ level, this linear convergence approach
only requires $6500$ tracking turns; when the angle $\xi$ is small,
fewer iterations are needed.

\subsection{Backward Tracking}

In the two recipes of section \ref{ssec:rec1} and \ref{ssec:rec2}
mentioned above, we need to find the set $C$ of $N+1$ backwards
tracked phase space points, and then launch spins at these points and
track forward so as to compute the set $B$.  In the case of linear
motion, it is trivial to obtain these backward tracked phase space
points.  One simply transforms $\vec z_0$ into the action--angle
variables of the linear motion and determines the phase advance per
turn of the linear motion.  Counting back these phase advances and
transforming the action--angle variables back into phase space leads
to the points $\vec c_j$. In the nonlinear case this would actually require
to track for $N$ more turns around the ring.

In the case of the linearly convergent method of section
\ref{ssec:rec2}, this extra effort becomes unnecessary.  One can start
with the phase space point $\vec z_0$ and launch three particles with
spins along $\vec e^{(1)}$, $\vec e^{(2)}$, and $\vec e^{(3)}$.
Tracking backward in azimuth defines the $N+1$ sets $P_j$ of the spins
$\vec p_j^{(1)}$, $\vec p_j^{(2)}$, and $\vec p_j^{(3)}$ with
\begin{equation}
  \vec p_j^{(k)} = \underline R^{-1}(\vec
  c_j)\cdot\ldots\cdot\underline R^{-1}(\vec c_1)\cdot\vec e^{(k)}\ .
\end{equation}
From the sets $P_j$ and $P_0$ one can again compute the spin transfer
matrix $\underline{\bar R}(\vec c_j)$ and with these matrices one
obtains the set $B$ with the elements $\vec b_j=\underline{\bar
  R}(\vec c_j)\cdot \vec n_0$, which again leads to $\vec n_N$ by
averaging.

\subsection{Forward Tracking}

There is an even more fundamental problem in the case of nonlinear
motion than the computation time.  When the lattice or the effect of
separate nonlinear elements are computed by nonlinear transfer maps,
the inverses of these maps might not be known.  In this case the
required phase space points $\vec c_j$ cannot be computed at all.
Nevertheless our method can be used to obtain the vector $\vec n_N$ as
follows. The arguments of the section on backward tracking can simply
be repeated for tracking forward.  One establishes the phase space
points $\vec c_j=\vec M(\vec c_{j-1})$ with $\vec c_0=\vec z_0$ for
$j\in\{1,\ldots,N\}$ and simultaneously the sets $S_j$ by tracking the
three unit vectors $\vec s_{j-1}^{(k)}$ for one turn with $\vec
s_0^{(k)}=\vec e^{(k)}$.  As in the fourth step of the recipe in
section \ref{ssec:rec2} one can then obtain the spin transfer matrix
$\underline{\bar R}(\vec c_j)$ from the phase space point $\vec z_0$
to $\vec c_j$.  The inverse of this transfer matrix is simply obtained
by transposition leading to the vectors $\vec b_j=\underline{\bar
  R}(\vec c_j)^T\cdot\vec n_0$.  The normalized average of the vectors
$\vec b_j$ is then the stroboscopic average $\vec n_{inv,N}$ of the
inverse motion.  For this average the error of the periodicity
condition for the inverse motion converges linearly to zero.
Fortunately, an $\vec n$-axis of the inverse motion $\vec n_{inv}(\vec
z)$ is also an $\vec n$-axis of the forward motion, since the
periodicity condition of the inverse motion reads as
\begin{equation}
  \underline R^{-1}(\vec M^{-1}(\vec z))\cdot\vec n_{inv}(\vec z) =
  \vec n_{inv}(\vec M^{-1}(\vec z))\ 
\end{equation}
and from this it follows that $\vec n_{inv}(\vec z)$ also obeys
the periodicity condition (\ref{eq:per}) of the forward motion.

\subsection{Tracking Rotation Angles and Rotation Vectors}

One can represent a rotation matrix by its rotation vector $\vec\beta$
and its rotation angle $\phi\in[0,\pi]$.  The simplest representation
is in terms of the vector $\vec\gamma=\sin(\phi/2)\vec\beta$ and
$\kappa=\cos(\phi/2)$.  The rotation matrix is then given by
\begin{equation}
R_{ij} = 2(\kappa^2\delta_{ij}+\gamma_i\gamma_j
-\kappa\varepsilon_{ijk}\gamma_k)-\delta_{ij}\ .
\end{equation}
This representation can be easily transported
through an accelerator by means of the equations
\begin{eqnarray}
  \tilde\kappa &=& \kappa_1\kappa_2 -\vec\gamma_1\cdot\vec\gamma_2 \ ,
  \ \ \kappa=|\tilde\kappa|\\ \vec\gamma&=&(\vec\gamma_1\kappa_2
  +\vec\gamma_2\kappa_1
  +\vec\gamma_2\times\vec\gamma_1)\frac{\tilde\kappa}{\kappa}\ .
\end{eqnarray}

It is therefore even more appropriate and faster by about a factor of
three to track rotation matrices represented by $\vec\gamma$ and
$\kappa$ directly, than to track three spin vectors.

\section{The Approximation of Linear Spin--Orbit Motion}

In this section we consider the special case of linear spin--orbit
motion as a way to illustrate and confirm the chief features of our
algorithm \cite{barberpc96}. For this purpose we define two periodic
vectors $\vec m$ and $\vec l$ orthogonal to $\vec n_0$ in order to
obtain an orthonormal right--handed dreibein at the previously
specified azimuth $\theta_0$.  We then write a spin vector $\vec s$ in
the form
\begin{equation}
  \vec s = \sqrt{1-\alpha^2-\beta^2}\vec n_0 + \alpha\vec m
  + \beta\vec l \ .
\label{eq:sab}
\end{equation}
In the case where both $\alpha$ and $\beta$ are much smaller than
unity we may treat the spin in a first order approximation in $\alpha$
and $\beta$ so that the spin vector (\ref{eq:sab}) can be written as
\begin{equation}
  \vec s = \vec n_0 + \alpha\vec m + \beta\vec l
\end{equation}
and is normalized in linear approximation. We combine the
two-component vector
\begin{equation}
  \vec{\tilde s}=\left(\begin{array}{c}\alpha\\ \beta\end{array}\right)
\end{equation}
and the orbit vector $\vec z$ into a single vector $\vec{\tilde z}$ of
d+2 components.  By linearizing the equations of motion with respect
to $\vec{\tilde z}$, an initial spin--orbit coordinate $\vec{\tilde
  z}_i$ at azimuth $\theta_0$ is mapped into a final coordinate
$\vec{\tilde z}_f$ after one turn around the ring \cite{chao81b} by
the $(d+2)\times (d+2)$ spin--orbit transfer matrix $\underline{\tilde
  M}$
\begin{equation}
  \vec{\tilde z}_f=\underline{\tilde M}\cdot\vec{\tilde z}_i \ ,\ \ 
  \underline{\tilde M} = \left(\begin{array}{cc} \underline
  M&\underline 0\\ \underline G&\underline D
\end{array}\right)\ ,
\label{eq:matrix}
\end{equation}
where $\underline M$ again denotes the $d\times d$ dimensional orbit
transfer matrix.  The matrices $\underline G$ and $\underline D$ are
$2\times d$ and $2\times 2$ matrices respectively with
\begin{equation}
  \vec{\tilde s}_f=\underline G\cdot\vec z_i+\underline
  D\cdot\vec{\tilde s}_i\ .
\end{equation}
In order to simplify the following formulas, we introduce the $2\times
d$ dimensional component $\underline G_m$ of the transfer matrix
$\underline{\tilde M}^m$ for $m$ turns around the ring.  We note that
the two component vector corresponding to $\vec n_0$ is $\vec{\tilde
  n}_0 = 0$ so that propagating this initial spin vector at a phase
space point $\vec z_i$ once around the ring leads to the final spin
$\vec{\tilde s}_f= \underline G_m\cdot\vec z_i$.  Therefore, we define
the two component vector $\vec{\tilde s}_N$ analogous to equation
(\ref{eq:sn}) by
\begin{equation}
  \vec{\tilde s}_N(\vec z_0)=\sum_{m=0}^N\underline G_m\cdot\underline
  M^{-m}\cdot\vec z_0\ , \ \ \underline{\tilde M}^m =
  \left(\begin{array}{cc} \underline M^m&\underline 0\\ \underline
    G_m&\underline D^m\end{array}\right)\ .\label{eq:sgm}
\end{equation}
From equation (\ref{eq:matrix}) it follows that
\begin{equation}
  \underline G_m\cdot\underline M^{-m}=\sum_{k=0}^{m-1}\underline
  D^k\cdot\underline G\cdot\underline M^{-(k+1)}\ .
\end{equation}
We expand the phase space coordinate $\vec z_0$ in eigenvectors $\vec
y_j$ of the one turn transfer matrix $\underline M$
\begin{equation}
  \vec z_0 = \sum_{j=1}^{d} a_j\vec y_j\ ,
\label{eq:eigen}
\end{equation}
with expansion constants $a_j$ and eigenvalues $\lambda_j^{-1}$
\begin{equation}
  \underline M\cdot\vec y_j = \lambda_j^{-1}\vec y_j \ .
\end{equation}
Thus we get
\begin{equation}
  \underline G_m\cdot\underline M^{-m}\cdot\vec y_j = \sum_{k=0}^{m-1}
  \lambda_j^{k+1}\underline D^k\cdot \underline G\cdot\vec y_j\ .
\label{eq:gmmm}
\end{equation}

We do not consider the case of spin--orbit resonances, so that the
matrices $\underline 1-\lambda_j\underline D$ are nonsingular. Thus
equation (\ref{eq:gmmm}) simplifies to

\begin{equation}
  \underline G_m\cdot\underline M^{-m}\cdot\vec y_j = \lambda_j
  (\underline 1-\lambda_j^m\underline D^m)\cdot (\underline
  1-\lambda_j\underline D)^{-1}\cdot\underline G\cdot \vec y_j \ .
\end{equation}
From this follows
\begin{equation}
  \sum_{m=0}^N\underline G_m\cdot \underline M^{-m}\cdot\vec y_j =
  \lambda_j[(N+1)\underline 1-(\underline 1-\lambda_j^{N+1}\underline
  D^{N+1})\cdot(\underline 1-\lambda_j\underline
  D)^{-1}]\cdot(\underline 1-\lambda_j\underline
  D)^{-1}\cdot\underline G\cdot\vec y_j\ ,
\end{equation}
so that
\begin{equation}
  \lim_{N\rightarrow \infty} [\frac{1}{N+1}\sum_{m=0}^N \underline
  G_m\cdot\underline M^{-m}\cdot\vec y_j] = \lambda_j(\underline
  1-\lambda_j\underline D)^{-1} \cdot\underline G\cdot\vec y_j\ .
\end{equation}
Combining this with equations (\ref{eq:sgm}) and (\ref{eq:eigen}) we get
\begin{equation}
  \vec{\tilde n}(\vec z_0) = \sum_{j=1}^d a_j\lambda_j(\underline
  1-\lambda_j \underline D)^{-1}\cdot\underline G\cdot\vec y_j \ ,
\end{equation}
where $\vec{\tilde n}$ is the two component vector corresponding to
$\vec n$. For $d=6$ this yields the well known expression of the $\vec
n$-axis for the case of linear spin--orbit motion
\cite{mane87b,barber94b}.  This confirms the method of stroboscopic
averaging for the linear case.  Also our predictions about convergence
speed are confirmed as follows.  In storage rings the particle motion
can only be stable if the eigenvalues have $|\lambda_j|\leq 1$. To
analyze the convergence speed, one has to realize that $\underline D$
is a rotation matrix and therefore that
\begin{eqnarray}
  |(\underline 1-\lambda_j^{N+1}\underline D^{N+1})\vec{\tilde s}|
  &\leq& (1+|\lambda_j|^{N+1})|\vec{\tilde s}|\leq 2|\vec{\tilde s}|\ 
  \ \ \ {\rm for\ all\ }\vec{\tilde s},\\ |\vec n_N(\vec z_0)- \vec
  n(\vec z_0)| &\leq& \frac{2}{N+1}\sum_{j=1}^d |a_j||(\underline
  1-\lambda_j\underline D)^{-2}\cdot\underline G\cdot\vec y_j|\ .
\end{eqnarray}
This inequality shows that also for linear spin--orbit motion one
finds convergence linear in $1/N$, and it can also be seen that the
convergence speed decreases with the orbital amplitudes $a_j$ and
becomes very slow close to first order spin--orbit resonances.

\section{Numerical Examples}

In order to illustrate which quantities can be computed and how
effective stroboscopic averaging can be in numerical computations, we
apply it to a model accelerator with known $\vec n$-axis.  Since this
model is somewhat artificial, we also apply our method to the vertical
motion of the HERA proton ring.  For simplicity, the vertical bends at
the interaction regions have been ignored.

\subsection{Comparison with an Analytically Solvable Model}

In this section we consider a special model \cite{mane88} with $d=2$
and $\vec{z}=(\Phi,J)$. The equations of motion are given by
\begin{eqnarray}
  \frac{d\Phi}{d\theta}(\theta) &=&\frac{Q}{2\pi} \ , \ \
  \frac{d J}{d\theta}(\theta)  =  0    \ , \nonumber\\
  \frac{d\vec s}{d\theta}(\theta) &=&
  \vec\Omega(\Phi(\theta),J )\times
  \vec s(\theta) \ ,
\end{eqnarray}
with
\begin{equation}
  \vec{\Omega}(\Phi,J) = \frac{1}{2\pi}(\nu_0\vec e_1 + \mu
  \sqrt{J} (\vec{e}_3 \sin \Phi + \vec{e}_2 \cos\Phi)) \ ,
\label{eq:OmM}
\end{equation}
where $\mu$ is a real parameter. Hence initial coordinates $\vec z_i$
are taken into final coordinates $\vec z_f$ by $\vec z_f=\vec M(\vec
z_i )$ with $\Phi_f=\Phi_i+ Q$ and $J_f=J_i$. In the following we
assume $\nu_0\neq Q$ and that $\nu_0/2\pi$ is not integer. This model
corresponds to the rotating field approximation often used to discuss
spin resonance in solid state physics \cite{abragam61}. We now
introduce the orthogonal matrix $\underline T(\vec e,\varphi)$
describing a rotation around a unit vector $\vec e$ by an angle
$\varphi$. Transforming the spin components of $\vec s$ into a
rotating frame by introducing $\vec s_{rot}=\underline T(\vec
e_1,\Phi)\cdot\vec s$, one obtains the simplified equation of spin
motion
\begin{equation}
\frac{d\vec s_{rot}}{d\theta}(\theta)  =
  \vec\Omega_{rot}(J)\times \vec s_{rot}(\theta) \ ,
\ \ \vec\Omega_{rot}(J) = \frac{1}{2\pi}\left(\begin{array}{c}
\nu_0-Q \\ \mu \sqrt{J} \\0 \end{array}\right) \ .
\end{equation}
If in this frame a spin field is oriented parallel to
$\vec{\Omega}_{rot}$, this field does not change from turn to turn.
Therefore $\vec n_{rot}=\vec{\Omega}_{rot}/|\vec{\Omega}_{rot}|$ is an
$\vec n$-axis.  In the original frame this $\vec n$-axis is
\begin{eqnarray}
  \vec n(\Phi,J) &=& \frac{\nu_0-Q}{|\nu_0-Q|}\frac{1}{\Lambda}
  ( (\nu_0-Q) \vec e_1 +\mu\sqrt{J} (\vec{e}_3 \sin \Phi
  +\vec{e}_2 \cos \Phi) ) \ ,\\ \Lambda &=& \sqrt{(\nu_0-Q)^2 +
    \mu^2 J} \ ,
\end{eqnarray}
where the `sign factor' $(\nu_0-Q)/|\nu_0-Q|$ has been chosen so that
on the closed orbit ($J=0$) the $\vec n$-axis $\vec n(\Phi,0)$
coincides with $\vec n_0 = \vec e_1$.  As with any function of phase
space this $\vec n$-axis is a periodic function in $\Phi$.

\begin{figure}[htbp]
\begin{center}
  \setlength{\unitlength}{1cm}
  \psfig{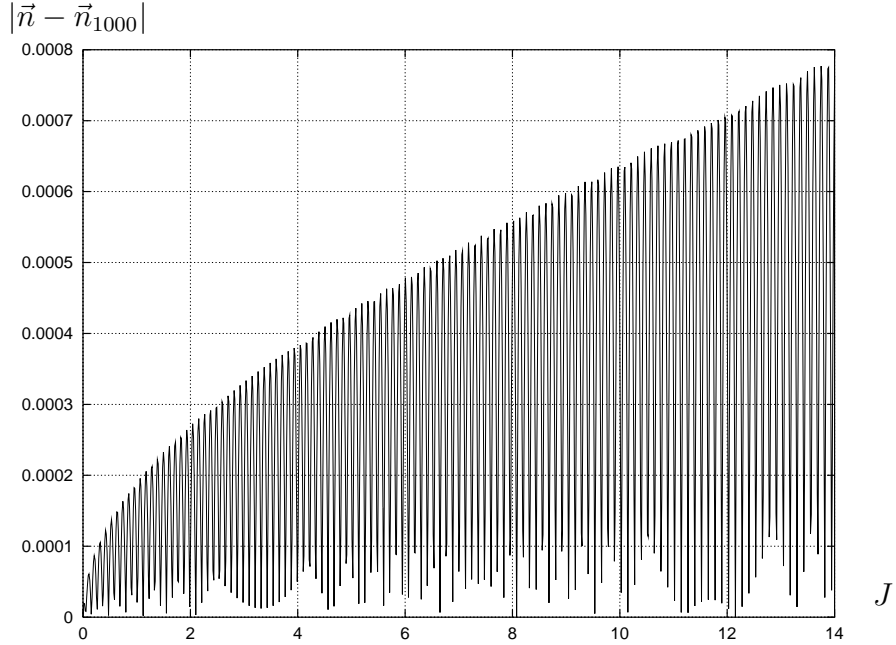}
\begin{picture}(0,0)(0,0)
  \put(-11.5,0.2) {$|\vec n-\vec n_{1000}|$}
  \put(0.0,-7.5) {$J$}
\end{picture}
\end{center}
\caption{The deviation of the stroboscopic average $\vec n_N$ from
  the analytically calculated $\vec n$ as a function of the amplitude
  $J$ in phase space}
\label{fg:modelconvary}
\end{figure}

\begin{figure}[htbp]
\begin{center}
  \setlength{\unitlength}{1cm}
  \psfig{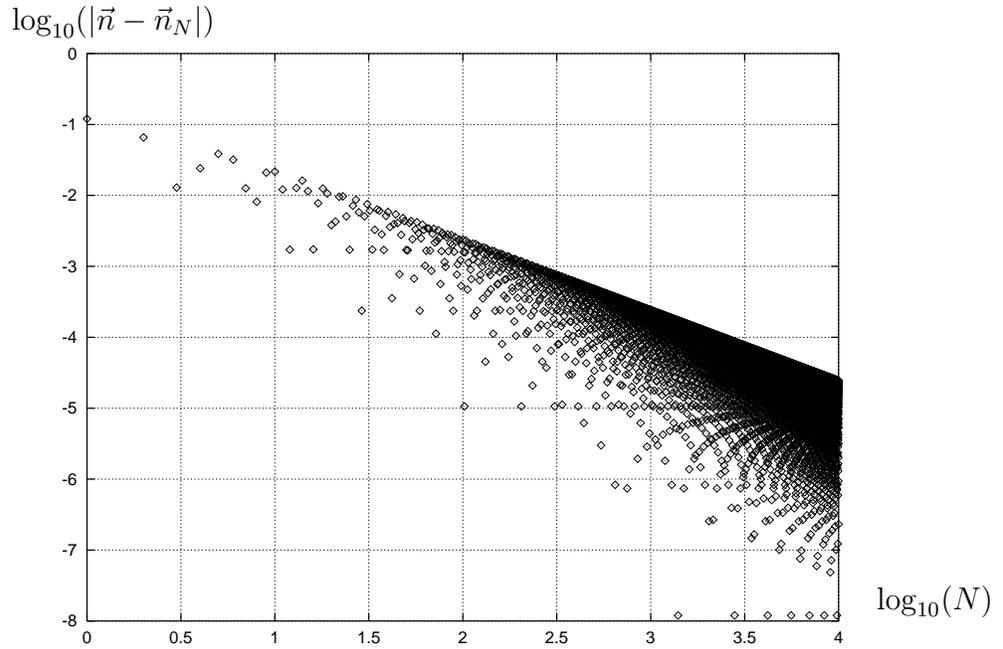}
\begin{picture}(0,0)(0,0)
  \put(-11.5,0.2) {$\log_{10}(|\vec n-\vec n_N|)$}
  \put(0.0,-7.5) {$\log_{10}(N)$}
\end{picture}
\end{center}
\caption{Convergence of the stroboscopic average $\vec n_N$ to
  the analytically calculated $\vec n$ with the number $N$ of turns
  tracked for $J=14$.}
\label{fg:modelconv}
\end{figure}

We now perform the stroboscopic average by the recipe of section
\ref{ssec:rec1} to compute an $\vec n$-axis at $\vec{z}_0=(\Phi,J)$
and $\theta_0$
\begin{eqnarray}
  \vec n_N&=&\frac{1}{N+1}\sum_{j=0}^N\prod_{k=1}^j\underline
  R(\Phi-kQ,J)\cdot\vec n_0\\ 
  &=&\frac{1}{N+1}\sum_{j=0}^N\prod_{k=1}^j\underline T(\vec
  e_1,-\Phi)\cdot\underline T(\vec n_{rot},\Lambda)\cdot\underline
  T(\vec e_1,\Phi)\cdot\vec n_0\\ &=&\frac{1}{N+1}\underline T(\vec
  e_1,-\Phi)\cdot\sum_{j=0}^N\underline T(\vec
  n_{rot},j\Lambda)\cdot\vec e_1\ .
\end{eqnarray}
If $\Lambda/2\pi$ is not an integer one obtains after some tedious
manipulations
\begin{eqnarray}
&& |\vec n_N(\Phi,J) -\vec n(\Phi,J)|
  =\sqrt{2}\sqrt{1-\tau_N} \ ,
\label{eq:errM} \\
&& \tau_N =  (  1 + \frac{\mu^2 J}{(N+1)^2(\nu_0-Q)^2}
\frac{1-\cos((N+1)\Lambda)}{1-\cos(\Lambda)})^{-1/2}\ .
\label{eq:tauN}
\end{eqnarray}
One sees that $|\vec n_N-\vec n|$ is an oscillating function of $J$
whose local maxima increase with $J$, reflecting the fact that large
orbital amplitudes reduce the convergence speed.  This behavior is
plotted in figure \ref{fg:modelconvary}. In this and the other figures
concerned with the solvable model we used the parameters $\nu_0=0.6\pi$,
$ Q=0.46\pi$, $\mu=0.2\pi$, and $\Phi=0.32$.  For large
$N$ equations (\ref{eq:errM}) and (\ref{eq:tauN}) predict that the
convergence is indeed linear in $1/N$, as illustrated by the
slope of $-1$ in figure \ref{fg:modelconv}.  Also one sees that
$|\vec n_N-\vec n|$ vanishes for $\Lambda/2\pi\neq$integer and if
$(N+1)\Lambda/2\pi=$integer. For $\Lambda/2\pi=$integer we have
\begin{equation}
\tau_N =(1+\frac{\mu^2 J}{(\nu_0-Q)^2})^{-1/2}\ .
\end{equation}
Therefore in this case $\vec n_N$ does not converge to $\vec n$.  This
is no surprise since the condition $\Lambda/2\pi=$integer amounts to
the resonance condition (\ref{eq:res}) which leads to this
non-uniqueness of the $\vec n$-axis.  It is interesting that the local
maxima of $|\vec n_N-\vec n|$ do not occur at these resonance points.

\begin{figure}[htbp]
\begin{center}
  \setlength{\unitlength}{1cm}
  \psfig{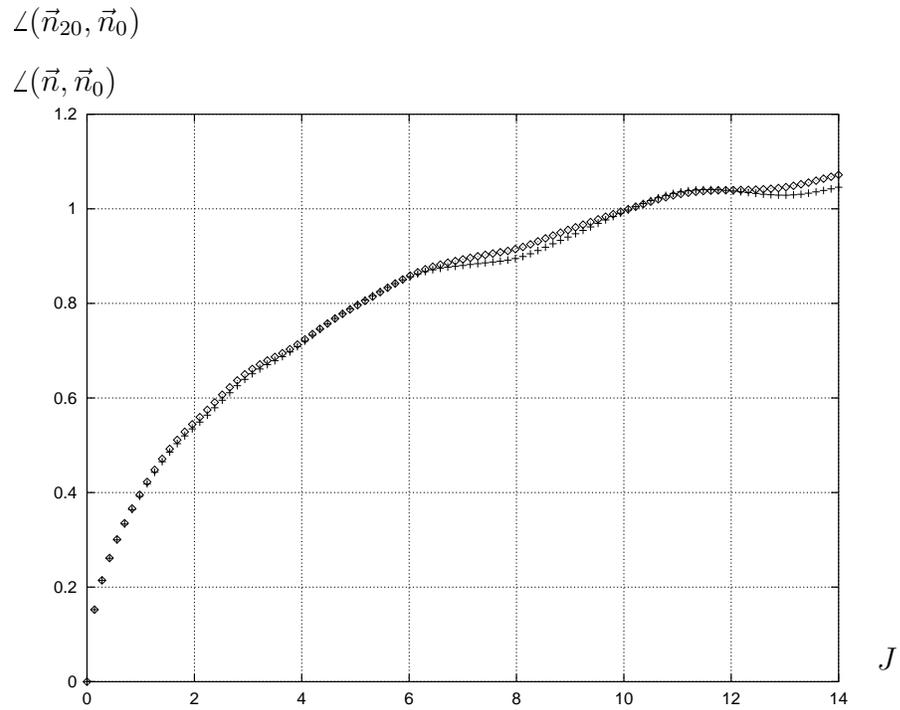}
\begin{picture}(0,0)(0,0)
  \put(-11.5,1.0) {$\angle(\vec n_{20},\vec n_0)$}
  \put(-11.5,0.2) {$\angle(\vec n,\vec n_0)$}
  \put(0.0,-7.5) {$J$}
\end{picture}
\end{center}
\caption{Opening angle of the analytically calculated
  $\vec n$ (diamonds) and the stroboscopic average $\vec n_N$ for
  $N=20$ (+) as a function of phase space amplitude $J$.}
\label{fg:modelangle}
\end{figure}

Figure \ref{fg:modelangle} shows the variation of the opening angle of
the polarization field as a function of orbital amplitude.  Since
$\vec n_0\cdot\vec n=|\nu_0-Q|/\Lambda$ as well as $\vec n_0\cdot\vec
n_N=\frac{1}{N+1}\sum_{j=0}^NT_{11}(\vec n_{rot},j\Lambda)$ do not
depend on the angle variable $\Phi$, the depicted angles are
equivalent to the phase averaged opening angle.  The good agreement
with the analytically computed opening angle of the $\vec n$-axis
shows that accurate values for the field can be obtained with a very
limited number of turns.

This analytically solvable model can also be used to illustrate the
construction of a phase independent spin rotation angle $\nu(J)$.
Having got an $\vec n$-axis, one can transform the spin components of
$\vec s$ into a rotated coordinate system ($\vec n,\vec u_1,\vec
u_2$).  With the simple choice
\begin{eqnarray}
  && \vec u_1(\Phi,J) = \frac{1}{|\vec n\times\vec e_1|} (\vec
  n\times\vec e_1)= \frac{\nu_0-Q}{|\nu_0-Q|} ( \vec{e}_2 \sin \Phi
  -\vec{e}_3 \cos \Phi) \ ,\\ && \vec u_2(\Phi,J) = \vec n\times\vec
  u_1 \ ,
\end{eqnarray}
both of which are clearly periodic in $\Phi$, one can show that the
spin rotation angle is independent of $\Phi$, namely
\begin{equation}
  \nu(J) = \frac{\nu_0-Q}{|\nu_0-Q|} \Lambda \ .
\label{eq:nuM1}
\end{equation}
However, $\nu(J)$ is free up to multiples of the orbit tune as
outlined in section \ref{ssec:conv}.  We can use this freedom to
obtain a $\nu(J)$ which reduces to $\nu_0$ on the closed orbit $(J=0)$.
The only choice which satisfies this condition is obtained by an
additional rotation of $\vec u_1$ and $\vec u_2$ around $\vec n$ by
$-\Phi$.
\begin{eqnarray}
&& \vec u_1(\Phi,J) = \vec{e}_1\frac{\mu\sqrt{J}}{\Lambda}\sin \Phi
 + \vec e_2\cos \Phi \sin \Phi (
 \frac{\nu_0-Q}{|\nu_0-Q|}-\frac{\nu_0-Q}{\Lambda}) \nonumber\\
&& -\vec e_3
           ( \frac{\nu_0-Q}{|\nu_0-Q|}\cos^2\Phi
  +\frac{\nu_0-Q}{\Lambda}\sin^2\Phi)  \ ,\\
&& \vec u_2(\Phi,J) = \vec n\times\vec  u_1  \ .
\end{eqnarray}
Both are again periodic in $\Phi$ and again we obtain a spin
rotation angle independent of $\Phi$, viz.
\begin{equation}
  \nu(J) = \frac{\nu_0-Q}{|\nu_0-Q|} \Lambda + Q \ .
\label{eq:nuM2}
\end{equation}
Correspondingly, on the closed orbit $\vec u_1,\vec u_2$ reduce to
\begin{equation}
\vec u_1(\Phi,0) = -\vec e_3 \frac{\nu_0-Q}{|\nu_0-Q|} \ ,\ \
\vec u_2(\Phi,0) =  \vec e_2 \frac{\nu_0-Q}{|\nu_0-Q|} \ .
\end{equation}

\subsection{Application to the HERA Proton Ring}

The HERA proton ring stores protons with an energy of $820$GeV.  At
this energy, the proton spin rotates about 1566.85 times around the
vertical direction in a flat ring during one turn.  The HERA ring is
not completely flat, nevertheless this number illustrates the
complexity of spin motion.  If a change of the phase space position of
a particle initiates a relative change of the one turn spin rotation
by only one part in $10^4$, the angle of precession has changed by
$56$ degrees.  This simple observation already hints at the strong
dependence of the equilibrium spin direction on the phase space
coordinates pointed out in the Introduction.  To illustrate this fact,
we restricted ourselves to vertical motion with a tune of 0.29,
ignored the vertical bends to obtain a flat ring, and computed the
third order expansion of the spin transfer matrix with respect to the
vertical phase space coordinates using the Differential Algebra code
COSY INFINITY \cite{cosyman6}.  Finally we stored the third order
phase space expansion of the corresponding rotation vector
$\vec\gamma$.  This power expansion was then used by the program
SPRINT to obtain the $\vec n$-axis.  Due to the strong phase space
dependence of spin motion, this expansion does not represent the real
spin motion in HERA very well, but it is nevertheless useful to
demonstrate the applicability of stroboscopic averaging.  We
transported spins by means of this rotation vector and approximated
the orbit motion by the linear transport matrix.

\begin{figure}[htbp]
\begin{center}
  \setlength{\unitlength}{1cm}
  \psfig{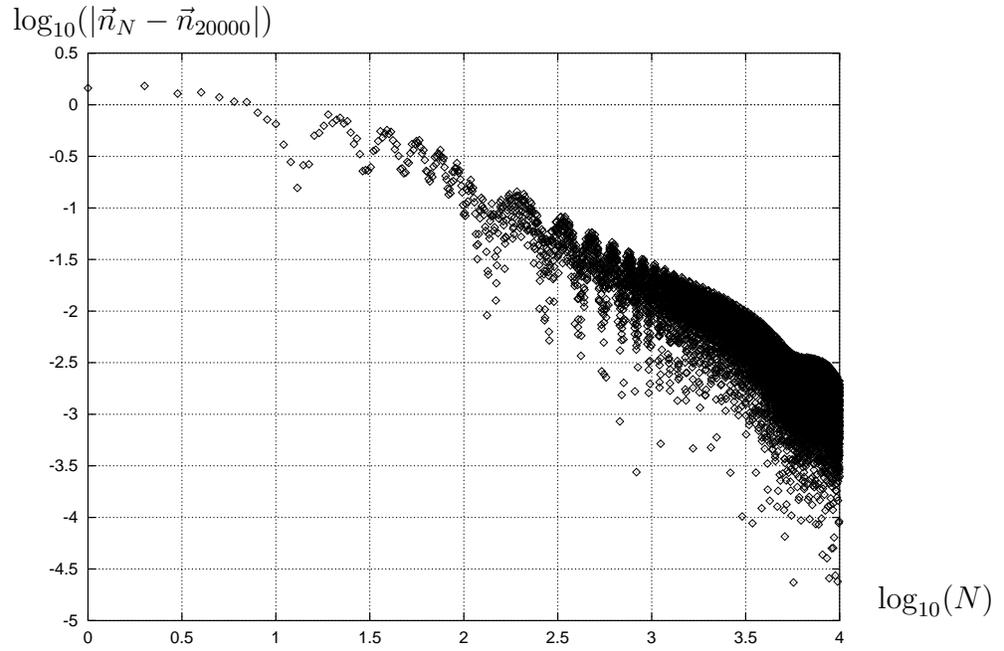}
\begin{picture}(0,0)(0,0)
  \put(-11.5,0.2) {$\log_{10}(|\vec n_N-\vec n_{20000}|)$}
  \put(0.0,-7.5) {$\log_{10}(N)$}
\end{picture}
\end{center}
\caption{Indication of Convergence}
\label{fg:heraconv}
\end{figure}

In a realistic accelerator, the existence of the $\vec n$-axis cannot
be guaranteed, but an approximately invariant spin field can be found,
if the series $\vec n_N$ converges. To indicate the convergence of
this series, we plot $|\vec n_N-\vec n_{20000}|$ for the phase space
point with $y_i=0.4$mm and $y'_i=0$ in the East interaction region.
This corresponds to an emittance of $69\pi$mm mrad and is therefore a
particle at approximately $4\sigma$ of the beam distribution.  The
slope of $-1$ in the double logarithmic scale of figure
\ref{fg:heraconv} illustrates clearly that the convergence is linear
in $1/N$ as derived in section \ref{ssec:conv}.

\begin{figure}[htbp]
\begin{center}
  \setlength{\unitlength}{1cm}
  \psfig{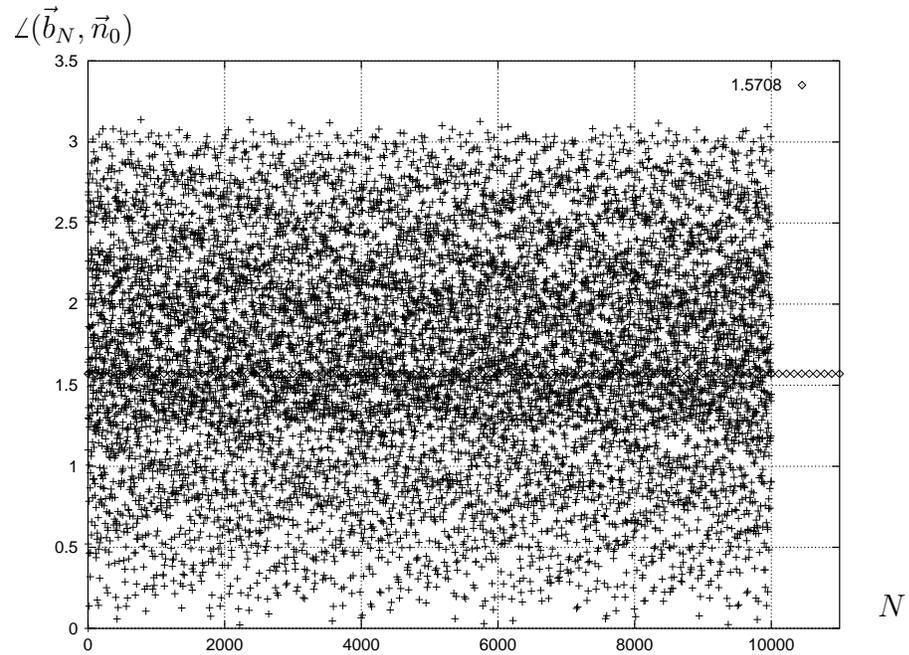}
\begin{picture}(0,0)(0,0)
  \put(-11.5,0.2) {$\angle(\vec b_N,\vec n_0)$}
  \put(0.0,-7.5) {$N$}
\end{picture}
\end{center}
\caption{Angle between the tracked spins and $\vec n_0$ for $y_i=0.4$mm.
  To guarantee convergence, this angle has to stay between 0 and
  $\pi/2$.}
\label{fg:heratrack}
\end{figure}

In that section, convergence could only be guaranteed if the angle
between the vectors $\vec b_N$ and $\vec n_0$ stayed smaller than
$\pi/2$ during tracking.  As an example we checked this requirement
for $y_i=0.4$mm and found that this condition is violated as illustrated
in figure \ref{fg:heratrack}, and convergence cannot be guaranteed ad
hoc.  Nevertheless, an approximately periodic spin field is obtained.

\begin{figure}[htbp]
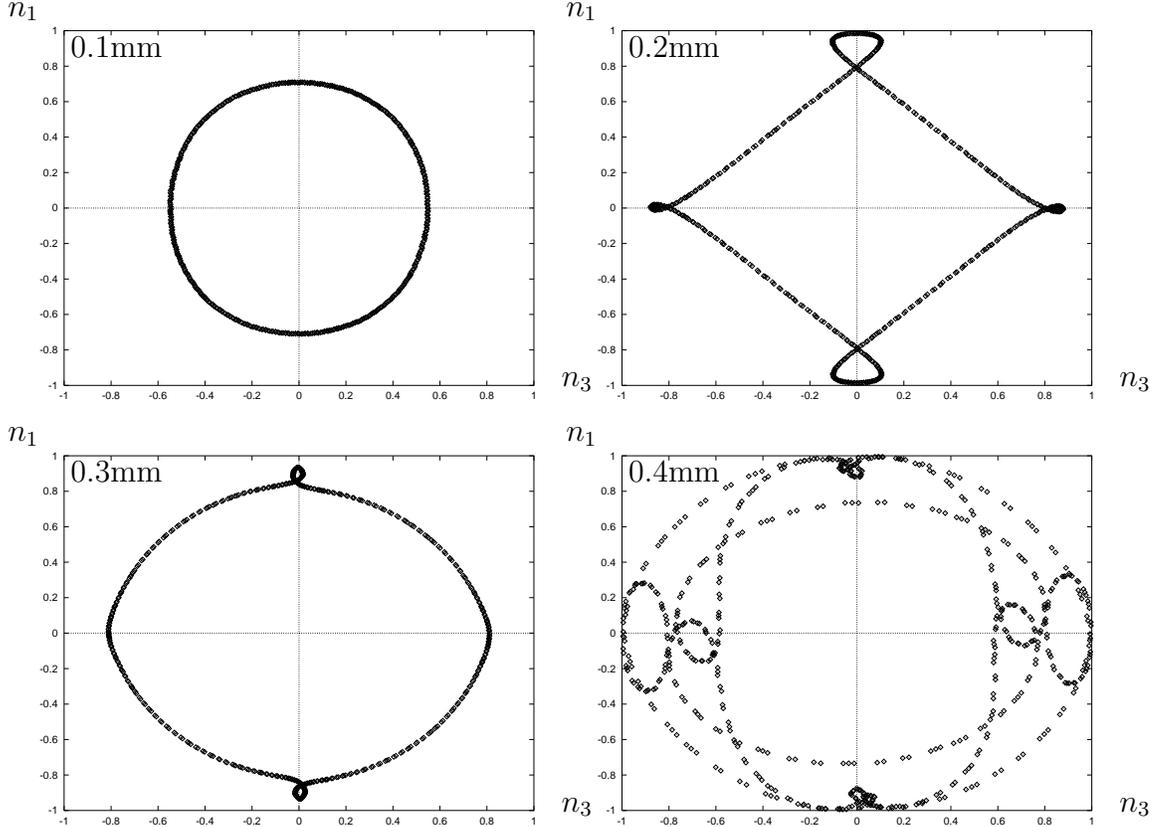

\begin{center}
  \setlength{\unitlength}{.64cm}
  \psfig{figure=hoffstatterpre7_1.inc,width=5cm,angle=-90}
\begin{picture}(0,0)(0,0)
  \put(-10.2,-0.65) {0.1mm}
  \put(-11.5,0.2) {$n_1$}
  \put(0.0,-7.5)  {$n_3$}
\end{picture}
  \psfig{figure=hoffstatterpre7_2.inc,width=5cm,angle=-90}
\begin{picture}(0,0)(0,0)
  \put(-10.2,-0.65) {0.2mm}
  \put(-11.5,0.2) {$n_1$}
  \put(0.0,-7.5)  {$n_3$}
\end{picture}

\vspace{2mm}

  \psfig{figure=hoffstatterpre7_3.inc,width=5cm,angle=-90}
\begin{picture}(0,0)(0,0)
  \put(-10.2,-0.65) {0.3mm}
  \put(-11.5,0.2) {$n_1$}
  \put(0.0,-7.5)  {$n_3$}
\end{picture}
  \psfig{figure=hoffstatterpre7_4.inc,width=5cm,angle=-90}
\begin{picture}(0,0)(0,0)
  \put(-10.2,-0.65) {0.4mm}
  \put(-11.5,0.2) {$n_1$}
  \put(0.0,-7.5)  {$n_3$}
\end{picture}
\end{center}
\caption{The stroboscopic average $\vec n_N$ for $N=12000$
  tracked for a further 600 turns for $y_i=$0.1mm, 0.2mm, 0.3mm, 0.4mm
  from top left to right bottom.}
\label{fg:heraphi1}
\end{figure}

One can easily check this by tracking a spin which is initially
parallel to $\vec n_N$ for many turns.  In order for $\vec n_N$ to
approximate an $\vec n$-axis, the tracked spins have to lie
approximately on a closed curve on the unit sphere.  Four such closed
curves were created by computing $\vec n_{12000}$ at the
phase space points $y_i=0.1$mm, $y_i=0.2$mm, $y_i=0.3$mm and
$y_i=0.4$mm with $y'_i=0$ and tracking for a further $600$ turns.  In
figure \ref{fg:heraphi1} we display the projections of these curves on
the horizontal plane.

\begin{figure}[htbp]
\begin{center}
  \setlength{\unitlength}{1cm}
  \psfig{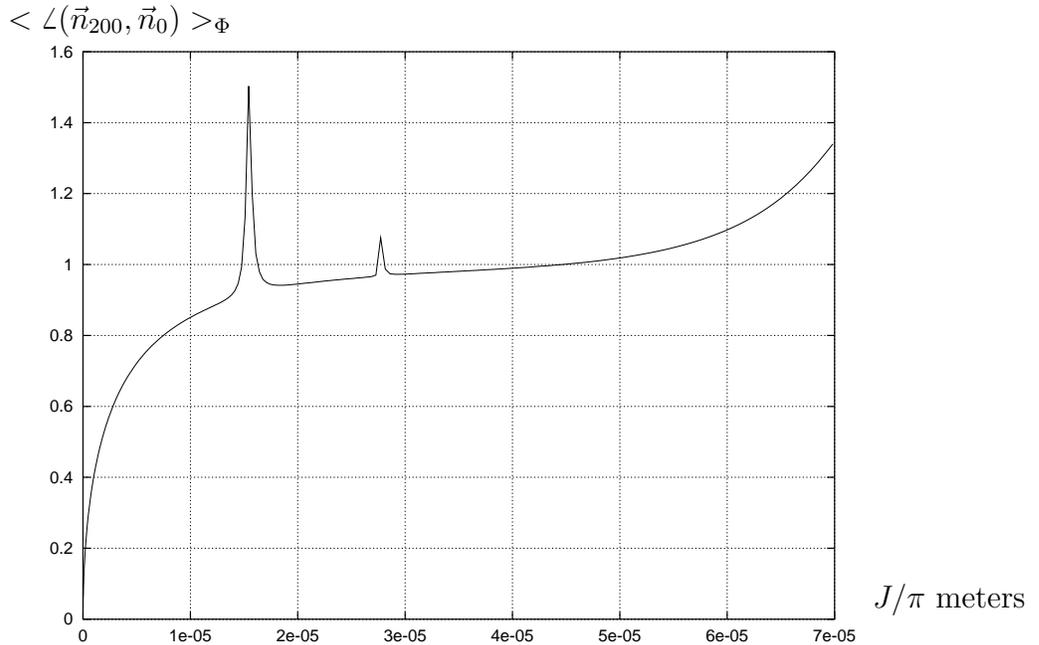}
\begin{picture}(0,0)(0,0)
  \put(-11.5,0.2) {$<\angle(\vec n_{200},\vec n_0)>_{\Phi}$}
  \put(0.0,-7.5) {$J/\pi$ meters}
\end{picture}
\end{center}
\caption{Phase averaged opening angle of the stroboscopic average
  $\vec n_N$ for $N=200$ as a function of the vertical phase space
  amplitude $J$.}
\label{fg:heraangle}
\end{figure}

Optimization of the average polarization of a particle beam requires
that the equilibrium polarization direction for every particle is
almost parallel to the average polarization direction. We therefore
averaged the angle between $\vec n_N$ and $\vec n_0$ over the orbital
phases and displayed this divergence for different phase space
amplitudes in figure \ref{fg:heraangle}.

\begin{figure}[htbp]
\begin{center}
  \setlength{\unitlength}{1cm}
  \psfig{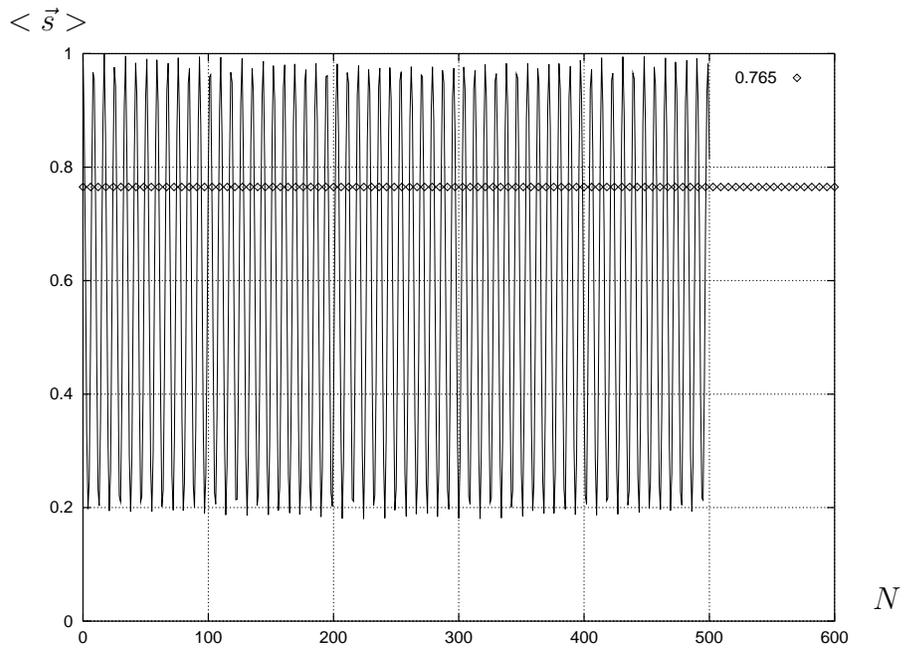}
\begin{picture}(0,0)(0,0)
  \put(-11.5,0.2) {$<\vec s>$}
  \put(0.0,-7.5) {$N$}
\end{picture}
\end{center}
\caption{Propagation of a beam that is initially completely polarized
  parallel to $\vec n_0$ leads to a fluctuating average polarization.
  For another beam that is initially polarized parallel to the
  periodic spin solution $\vec n$ the average polarization stays
  constant, in this case equal to 0.765.}
\label{fg:heraparal}
\end{figure}

Yet another way of illustrating the importance of the $\vec n$-axis is
illustrated in figure \ref{fg:heraparal}.  Particles at 100 different
phases at a normalized emittance of about $4\pi$mm mrad, corresponding
to $y_i=0.1mm$ and $y'_i=0$, have been tracked through HERA for 500
turns while the beam was initially polarized 100\% parallel to $\vec
n_0$.  Similar kinds of tracking results have been presented in
\cite{balandin96}.  Since this polarization distribution is not the
equilibrium distribution, the averaged polarization exhibits a strong
beat.  This figure also shows that when spins at phase space
coordinate $\vec z$ are initially parallel to $\vec n(\vec z)$, the
averaged polarization stays constant.  Therefore, by starting
simulations with spins parallel to the $\vec n$-axis one can perform a
much cleaner analysis of beam polarization in accelerators.

\section*{Conclusion}
\addcontentsline{toc}{section}{Conclusion}

We have introduced an algorithm for computing the
Derbenev--Kondratenko spin axis ($\vec n$-axis) from straight forward
spin phase space tracking data with the following features:
\begin{itemize}
\item It can be implemented in any existing spin tracking program.
\item For an accuracy on the $10^{-3}$ level typically less than 3000
  turns have to be tracked.
\item No artificial damping is needed.
\item Since our method is non-perturbative, no resonance factors appear
  in the algorithm and it is applicable even near spin--orbit
  resonances.
\end{itemize}
With the method of stroboscopic averaging important features of
accelerators can now be analyzed.  One very significant field of study
will be the computation of spin tune spreads, which previously could
hardly be analyzed and are now easily accessible.  Diffusion and
damping terms could be added to the analysis and $\gamma\partial\vec
n/\partial\gamma$ \cite{yokoya92} can be computed for the simulation
of electron polarization in storage rings.  In addition numerical
checks of uniqueness, convergence near resonances, and the effect of
nonlinear orbit motion should be made.

It is the opinion of the authors that now, after the introduction of
stroboscopic averaging, spin tracking in storage rings should always
be initialized with spins parallel to the equilibrium distribution,
since much clearer analysis becomes possible.

\section*{Acknowledgments}
\addcontentsline{toc}{section}{Acknowledgments}

We wish to thank Desmond P.\ Barber for careful reading of and valuable
remarks on the manuscript and Gerhard Ripken for useful discussions.


\end{document}